\documentclass[twocolumn,showpacs,preprintnumbers,amsmath,amssymb]{revtex4}

\usepackage{graphicx}
\usepackage{dcolumn}
\usepackage{bm}

\usepackage{color}

\usepackage{amssymb}

\usepackage{epsfig,enumerate}

\newcommand{\up}{\uparrow}
\newcommand{\dn}{\downarrow}

\newcommand{\qE}{{\mathfrak E}}

\newcommand{\vacv}{v}

\newcommand{\texp}{\stackrel{\leftarrow}{\exp}}
\newcommand{\K}{\mathbf K}

\newcommand{\C}                 {{\bf C}}

\newcommand{\la}{{\langle}}
\newcommand{\ra}{{\rangle}}

\newcommand{\cD}{{\cal D}}

\newcommand{\cH}{{\cal H}}

\newcommand{\bP}{{\bf P}}

\newcommand{\bu}{{\bf u}}

\newcommand{\bU}{{\bf U}}

\newcommand{\cL}{{\cal L}}

\newcommand{\ds}{\displaystyle}

\newcommand{\be}{\begin{equation}}
\newcommand{\ee}{\end{equation}}
\newcommand{\ba}{\begin{array}}
\newcommand{\ea}{\end{array}}

\newcommand{\tr}{{\rm tr}}
\newcommand{\er}[1]{\hbox{(\ref{#1})}}

\newtheorem{theorem}            {Theorem}[section]

\newtheorem{sideremark}         [theorem]{Remark}
\newtheorem{sideeg}           [theorem]{Example}
\newtheorem{sideconj}           [theorem]{Conjecture}
\newtheorem{sideassumption}   [theorem]{Assumption}

\newcommand{\R}                 {{\bf R}}

\newcommand{\E}                 {{\bf E}}

\newcommand{\inverse}[1]        {{\textstyle\frac{1}{#1}}}
\newcommand{\demi}              {\inverse{2}}

\date{December 13, 2004 \ Revised March 22, 2005}

\begin{document}           

\title{A Quantum Langevin Formulation of Risk-Sensitive Optimal
Control\thanks{This work was supported by the Australian Research
Council.}}

\author{M.R.~James\thanks{Department of Engineering, Australian National University,
Canberra, ACT 0200,  Australia. Matthew.James@anu.edu.au }
 }



\begin{abstract}
In this paper we formulate a risk-sensitive optimal control problem for continuously monitored open quantum systems modelled by quantum Langevin equations. The optimal controller
is expressed in terms of a modified conditional state, which we call a risk-sensitive state, that represents measurement knowledge tempered by the control purpose.  One of the two components of the optimal controller is dynamic, a filter that computes the risk-sensitive state.
 The second component is an optimal control feedback function that is found by solving the dynamic programming equation. The optimal controller can be implemented using classical electronics.  
 The ideas are illustrated using an example of feedback control of a two-level atom.
\end{abstract}


\maketitle


\section{Introduction}
\label{sec:intro}

Recent years have seen significant advances in quantum technology, quantum information and computing, continuous quantum measurements, and feedback control is playing an increasing role, for example \cite{GM96}, \cite{HC93}, \cite{WM93}, \cite{WM93a}, \cite{WM95}, \cite{NC00}, \cite{AASDM02},  \cite{VPB92}, \cite{VPB92a}, \cite{BGM04}, \cite{VPB83}, \cite{DJ99}, \cite{HSM04}, \cite{BEB04}, \cite{J04}.  Optimal feedback control is an important methodology from classical control theory that is widely used  for control system design, and has been applied to quantum systems, for example \cite{VPB83}, \cite{VPB88}, \cite{DJ99}, \cite{BEB04}, \cite{J04}. The designer encodes the desired performance objectives (e.g. regulation,  etc) in a cost function, which is then optimized. When the system to be controlled is subject to noise, uncertainty, or disturbances, optimal feedback  controllers are sought. Because of the intrinsic randomness in quantum measurements, feedback control of quantum systems using classical electronics has close connections to classical stochastic control theory, \cite{DHJMT00}.

In the context of classical linear systems subject to Gaussian noise, there are two main types of cost functions. In linear quadratic gaussian (LQG) design, the cost is an expected value of an integral or sum of quadratic system variables; the cost is additive, \cite{KV86}. In contrast, in linear exponential quadratic gaussian (LEQG) control, the cost is multiplicative, being the average of the exponential of an integral or sum, \cite{J73}, \cite{W81}.  LEQG is also known as risk-sensitive optimal control (LQG is sometimes referred to as risk-neutral optimal control). More general formulations of these problems have been developed, e.g. for nonlinear stochastic systems. It is known that risk-sensitive controllers enjoy enhanced robustness properties, \cite{DGKF89}, \cite{DJP00}, and this provides an important motivation for the study of risk-sensitive problems. A fundamental difference between the optimal solutions to the risk-neutral and risk-sensitive problems is that the optimal risk-neutral controller is given in terms of the optimal state estimator (Kalman filter in the LQG case), while the optimal risk-sensitive controller is given in terms of a quantity that takes into account the cost objective (it is given by a modified Kalman filter in the LEQG case), and is in general not the optimal state estimator, \cite{W81}, \cite{BV85}.

Optimal feedback control problems for quantum systems using additive cost functions have been considered in the literature,  \cite{VPB83}, \cite{VPB88}, \cite{DJ99}, \cite{BEB04}; we refer to these as risk-neutral. The optimal risk-neutral controllers obtained in these papers are given in terms of  a posteriori conditional states (evolving according to  stochastic master equations, or quantum trajectory equation, or Belavkin quantum filtering equations). In \cite{J04} a class of risk-sensitive optimal control problems was considered.  This class of problems was specified in discrete-time using the framework of quantum operations and conditional states. The optimal risk-sensitive controller is given in terms of a modified unnormalized conditional state that takes into account the cost function.  This risk-sensitive state represents measurement knowledge available to the controller tempered by the purpose of the designer, as in the classical case.

In this paper we formulate a risk-sensitive optimal control problem for continuously monitored open quantum systems. We use a Markovian approximate model to describe the open quantum system, \cite[Chapters 5 and 11]{GZ00}. This model is given by a quantum Langevin equation.  Quantum stochastic calculus and dynamic programing methods are used to study the optimal control problem. A heuristic  solution is given (as in classical continuous time measurement feedback stochastic optimal control, there are substantial technical issues). As in \cite{J04}, the solution is given in terms of a modified or risk-sensitive conditional state, and we present the corresponding modified stochastic master equation, a risk-sensitive quantum filter. We also consider briefly a risk-neutral problem, also formulated using quantum Langevin equations, to facilitate connection with the results in the papers \cite{VPB83}, \cite{VPB88},  \cite{DJ99}, \cite{BEB04} via quantum filtering.

We begin in Section \ref{sec:prob} by formulating the risk-sensitive problem. Then in Section \ref{sec:rep} we show how the cost function can be expressed as a stochastic representation in terms of the risk-sensitive state mentioned above. The dynamic programming solution is discussed in Section \ref{sec:dp}, and the risk-neutral problem is summarized in Section \ref{sec:rn}.  In Section \ref{sec:2l} we illustrate our results in the context of feedback control of a two-level atom, \cite{BEB04}.

\section{Problem Formulation}
\label{sec:prob}


We consider the problem of controlling an open quantum system
model with the following features:
\begin{enumerate}
\item
The evolution can be influenced by control variables $u$ that
enter the system Hamiltonian $H(u)$.
\item
The system $S$ interacts with two heat baths (electromagnetic
field channels) $B_1$ and $B_2$.
\item
Channel $B_1$ is not monitored; its influence may be used, for example,  to model dissipative effects.

\item
Channel $B_2$ is continuously monitored, providing weak measurements
of the system $S$, the results $y$ of which are available to the
{\em controller} $\K$, a classical system which processes this information to produce the control actions $u$.

\item
The control in general is allowed to be a causal function of the
measurement trajectory (not just a function of the current
measurement value).

\item
The controller $\K$ is chosen so that it minimizes a suitable cost
or performance function $J(\K)$.

\end{enumerate}

The controls $u$ take values in a set $\bU$, say real or complex vectors of dimension $m$ ($\bU=\R^m$ or $\bU=\C^m$). In general the set $\bU$ may be bounded or unbounded. Of course, we can have multiple measured and unmeasured field channels, though we use one of each for notational simplicity.

We now describe the ideal dynamics of the controlled system using quantum stochastic differential equations, \cite{GZ00}, \cite{KRP92}. Let 
$u=u(\cdot)$ be a control signal (a function of time $t$ with values $u(t)\in\bU$). Consider the interaction picture
unitary operators $U(t)=U^u(t)$ (often we omit explicit dependencies on $u$ from the notation) solving the quantum stochastic
differential equation (QSDE) \cite[eq. (11.2.7)]{GZ00}, \cite[sec. 26]{KRP92},
\begin{eqnarray}
dU(t)&=& \{ -K(u(t)) dt + L dB_1^\dagger(t) - L^\dagger dB_1(t)
\nonumber \\
&& \hspace{0.5cm}
+ M dB_2^\dagger(t) - M^\dagger dB_2(t)
\} U(t)
\label{U-qsde}
\end{eqnarray}
with initial condition $U(0)=I$, where
\be
K(u) = \frac{i}{\hbar} H(u) +\frac{1}{2} L^\dagger L +\frac{1}{2}
M^\dagger M .
\ee
Here, $L$ and $M$ are system operators which together with the
field operators $b_1(t)=\dot B_1(t)$, $b_2(t)=\dot B_2(t)$, model
the interaction of the system with the channels. Note that equation
\er{U-qsde} is written in Ito form (see, e.g. \cite[Chapter 4]{CWG04}), as will all
stochastic differential equations in this paper. With vacuum initialization of the field channels, the two non-zero Ito products are, \cite[eq. (11.2.6)]{GZ00},
\be
dB_1(t) dB_1^\dagger(t)=dt \ \text{and} \ dB_2(t) dB_2^\dagger(t)=dt .
\ee

Then system
operators $X$ evolve according to\footnote{The notation $j_t(X)$ is used for  the family of solution operators for the quantum Langevin equation, \cite[Sec. 28]{KRP92}.}
\be
X(t) = j_t(u, X) = U^\dagger (t) X U(t)
\ee
and satisfy the quantum Langevin equation (QLE)
\begin{eqnarray}
dX(t) &=& (-X(t) K(t) - K^\dagger(t) X(t)  
\label{X-qle} \\
&&+ L^\dagger(t) X(t) L(t) + M^\dagger(t) X(t) M(t) 
)dt 
\nonumber
\\
&&+ [ X(t),L(t)] dB_1^\dagger(t) - [X(t),L^\dagger(t)] dB_1(t) 
\nonumber \\
&&+  [ X(t),M(t)] dB_2^\dagger(t) - [X(t),M^\dagger(t)] dB_2(t)
\nonumber 
\end{eqnarray}
where $L(t) = j_t(u,L)$, $M(t) = j_t(u,M)$, and $K(t) = j_t(u, K(u(t)))$ (note the slight abuse of notation regarding $K(\cdot)$).

 In terms of states, if $\pi_0$ is a given system state, we write  
\be
\rho_0 =  \pi_0 \otimes \vacv_1\vacv^\dagger_1 \otimes
\vacv_2\vacv_2^\dagger
\ee
 (the channels are initially in their vacuum states $v_1$ and $v_2$ respectively), and so the state of the system plus channels  at time $t$ is given by
\be
\rho(t) = U(t) \rho_0 U^\dagger(t) ,
\ee
so that
\be
\la \rho_0, j_t(u,X) \ra = \la \rho(t), X\otimes I \otimes I   \ra .
\ee
Here, we have used the notation
\be
\la A,B \ra = \tr[A^\dagger B] ,
\ee
and the symbol $I$ denotes the appropriate identity operator.
When $u(\cdot)$ is an open loop signal, or simply constant  (no feedback), we can trace out the field channels and obtain the master equation for our setup. Indeed, if $\bar \rho(t)$ denotes the partial trace of $\rho(t)$ obtained by tracing out the field channels, then $\bar \rho(t)$ solves the master equation
\begin{eqnarray}
\dot{\bar \rho}(t) &=& - K(u(t)) \bar \rho(t) - \bar \rho(t) K^\dagger(u(t))
\nonumber \\
&& \hspace{0.5cm} + L\bar \rho(t) L^\dagger + M\bar \rho(t) M^\dagger 
\nonumber 
\\
&=& - \frac{i}{\hbar} [H(u(t)),\bar \rho(t) ] + \cD[L] \bar \rho(t) + \cD[M]\bar \rho(t) ,
\label{master}
\end{eqnarray}
where $\cD[c]\rho = c\rho c^\dagger -\demi (c^\dagger c \rho + \rho c^\dagger c)$ is the decoherence operator. The initial condition for \er{master} is $\bar\rho(0)=\pi_0$.

We regard the field operators $B_k(t)$, $k=1,2$, as input fields
\cite[Section 11.3.2]{GZ00}, with corresponding output fields
$A_k(t)$ defined by
\be
A_k(t) = j_t(u,B_k(t)) .
\ee
The real quadratures of the input fields are defined by
\be
Q_k(t) = B_k(t) + B^\dagger_k(t) ,
\ee
and we write
\be
Q(t)= \left( \ba{c} Q_1(t) \\   Q_2(t) \ea \right) ,
\ee
a vector of independent quantum noises. For the output field real quadratures we write
\be
Y_1(t) = j_t(u,Q_1(t)), \ \text{and} \ Y_2(t) = j_t(u,Q_2(t)).
\ee
These processes satisfy the QSDEs
\begin{eqnarray}
dY_1(t) &=& (L(t)+L^\dagger(t))dt + dQ_1(t)
\nonumber \\
dY_2(t) &=& (M(t)+M^\dagger(t))dt + dQ_2(t) .
\label{Y-qsde}
\end{eqnarray}

We continuously monitor the second channel, and measurement of $Y_2(t)$ produces a real output  measurement signal
$y_2(t)$, which is used by a (classical) controller $\K$ to produce
the input control signal $u(t)$ by
\be
u(t) = \K(t, y_{2,[0,t]}) .
\label{u-K}
\ee
The notation used in \er{u-K} is meant to indicate the causal
dependence of the control on the measurements; $y_{2,[0,t]}$
indicates the segment of the measurement signal on the time
interval $[0,t]$, so in effect the controller $\K = \{ \K(t,
\cdot)
\}$ is a family of functions. To illustrate, consider the following two special cases: (i) static feedback, where $u(t)=k(y(t))$, so that the control at time $t$ depends only on
the measurement at time $t$; (ii) dynamic feedback, where $u$ is determined by
\be
\K \ : \ \ba{rl}
d \zeta(t) & = f_\K (\zeta(t)) dt + g_\K(\zeta(t))dy_2(t)
\\
u(t) & = h_\K (\zeta(t),y_2(t)) .
\ea
\label{K-dyn}
\ee 

The complete dynamics of the open system and controller is
obtained by combining the open system evolution
\er{U-qsde} or the QLE \er{X-qle} with the control law \er{u-K}. Equations \er{U-qsde} and \er{X-qle} continue to hold since $u$ depends causally on $y_2$, \cite{VPB88}.

We now specify the cost function that we will use to determine the
\lq\lq{best}\rq\rq \ choice of controller. It will be defined over a fixed time interval $[0,T]$.
Let $C_1(u)$ be a non-negative self-adjoint system operator
depending on the control value $u$, and let $C_2$ be a
non-negative self-adjoint system operator. These so-called {\em
cost operators} are chosen to reflect the performance objectives,
and explicitly include the control so that a balance between
performance and control cost can be achieved (see Section \ref{sec:2l} for an example).
The quantity
\be
\int_0^T C_1(t) dt + C_2(T),
\label{run}
\ee
where $C_1(t) = j_t(u,C_1(u(t)))$, $C_2(t)=j_t(u,C_2)$, accumulates
cost over the given  time interval and provides a penalty for the final
time (we again take the liberty of a slight abuse of notation). Instead of using the expected value of the quantity
\er{run} as a cost function (risk-neutral case, see Section \ref{sec:rn}),  we consider the average of the
exponential of \er{run} in the following way. Define $R(t)$ to be
the solution of the operator differential equation
\be
\frac{dR(t)}{dt} = \frac{\mu}{2}C_1(t) R(t)
\label{R-rs}
\ee
with initial condition $R(0)=I$. Here, $\mu > 0$ is a positive
(risk) parameter. The solution of \er{R-rs} can be expressed as
the time-ordered exponential
\be
R(t) = \texp \left(\frac{\mu}{2} \int_0^t C_1(s) ds \right) .
\ee
We then define the {\em risk-sensitive} cost function to be the quantum
expectation
\be
J^\mu(\K) = \la \rho_0, R^\dagger(T) e^{\mu C_2(T)} R(T) \ra .
\label{J-rs}
\ee
Here, $\rho_0 =  \pi_0 \otimes \vacv_1\vacv^\dagger_1 \otimes
\vacv_2\vacv^\dagger_2$, as above. 

The cost function \er{J-rs} is one possible quantum generalization of the classical risk-sensitive criterion. The operator ordering was chosen to be compatible with the evolution of operators in the Heisenberg picture, thus facilitating the stochastic dynamical representations given in Section \ref{sec:rep} which are needed for dynamic programming solution in Section \ref{sec:dp}.




\section{Stochastic Representation of Cost and Filtering}
\label{sec:rep}

We wish to express the risk-sensitive cost \er{J-rs} in terms of a (reduced) quantity defined on the system space that is driven by the data $y_2(t)$, $0 \leq t \leq T$, obtained from the continuous monitoring of the second field channel. To this end we first represent the cost $J^\mu(\K)$ as a classical expectation with respect to a reference Wiener distribution in terms of (reduced) quantities defined on the system space, analogous to the stochastic representations of quantum dynamical semigroups considered in  \cite{VPB92}, \cite{AH01}, and then we apply classical filtering. Essentially, we monitor both channels and then average the results for the first channel. This procedure does not \lq\lq{demolish}\rq\rq \ system variables due to the commutativity property or {\em quantum non-demolition} (QND) condition:
for all initial system operators $X$:
\be
[Q(t),X]=0 \ \forall \ 0 \leq t \leq T ;
\label{bqnd-1}
\ee
see  \cite[eq. (5.3.29)]{GZ00}, \cite{VPB92}, \cite{VPB92a} (equation \er{bqnd-1} is understood componentwise).

Before considering the cost representation, we discuss the statistics of the  fields, \cite[Chapters 5 and 11]{GZ00}. The input real quadrature operators $Q(t),  0 \leq t \leq T$, are commutative,
\be
[Q_j(t),Q_k(s)]=0 \ \forall \ 0 \leq s, t \leq T, \ j,k=1,2,
\ee
and when the fields are initialized in the vacuum states they correspond to a (two-dimensional) real Wiener process (Brownian motion) $q(t)=(q_1(t),q_2(t))$ via the Segal map \cite[Chapter 5]{AH01}.  Indeed, let $\Omega_T$ denote the set of all Wiener paths (continuous functions of time on the interval $[0,T]$). The probability of a subset $F\subset\Omega_T$ of paths is 
\be
\bP^0(F) = \la  \vacv_1\vacv^\dagger_1 \otimes
\vacv_2\vacv^\dagger_2, P^Q_T(F) \ra ,
\label{P0}
\ee
where $P^Q_T(F)$ is the projection operator associated with $F$ and $Q(s)$, $0 \leq s \leq T$. The probability distribution $\bP^0$ is the Wiener distribution, under which the increments $q(t)-q(s)$, $0 \leq s \leq t \leq T$, are independent, Gaussian, with zero mean and covariance $(t-s)I$ (here $I$ is the $2\times 2$ identity matrix). The output fields $Y(t)=j_t(u,Q(t))$ are also commutative
\be
[Y_j(t), Y_k(s)]=0 \ \forall \ 0 \leq s, t \leq T, j,k=1,2,
\label{bqnd-y}
\ee
and satisfy the QND condition (cf. \er{bqnd-1})
\be
[ Y(s),X(t)]=0 \ \forall \ 0 \leq s \leq t \leq T,
\label{bqnd-2}
\ee
see \cite[Eq. (2.24)]{BS92},
\cite[Eq. (8)]{VPB92}, \cite[Section 5.3]{GZ00}. The statistics of the continuously observed channel is discussed below (see \er{output-prob}).

We continue now with the representation of the cost.  Define an operator $V(t)$, not unitary in general, by
\be
V(t)=U(t)R(t) .
\ee
Then by the rules of quantum stochastic calculus \cite[Section 5.3]{GZ00}, \cite[Chapter III]{KRP92}, \cite[Chapter 5]{AH01} $V(t)$ solves the QSDE
\begin{eqnarray}
dV(t)&=& \{ -K^\mu(u(t)) dt + L dB_1^\dagger(t) - L^\dagger dB_1(t)
\nonumber \\
&&\hspace{0.5cm} + M dB_2^\dagger(t) - M^\dagger dB_2(t)
\} V(t)
\label{V-qsde}
\end{eqnarray}
with initial condition $V(0)=I$, where
\be
K^\mu(u)= K(u) - \mu \demi C_1(u) .
\ee
If we define
\be
j^\mu_t(u,X)= V^\dagger(t) [X \otimes I \otimes I ]  V(t) ,
\ee
then we can write
\be
J^\mu(\K) =  \la \rho_0,   j^\mu_T(u,e^{\mu C_2}) \ra .
\ee

We now apply the technique developed in \cite[Section 2]{VPB92} and described in \cite[Chapter 5]{AH01} for stochastic representation of quantum semigroups $\Phi_t(X)=\qE_{v_1\otimes v_2}[j_t(u,X)]$\footnote{$\qE_{v_1\otimes v_2}$ denotes conditional expectation with respect to the vacuum states, \cite[Chapter 5]{AH01}.}
 to obtain a stochastic representation of the cost. The key idea is that, with vacuum initialization, $V(t)$ can be viewed as a function of the real quadratures of the fields, and by the Segal duality map (see \cite[Section 5.2.1]{AH01}), this means that quantum expectations are equivalent to classical expectations. 
 The first point can be seen from the fact that $dB_1(t)v_1=0$, which means that we can write
\be
 (L dB_1^\dagger(t) - L^\dagger dB_1(t)) v_1 =  L dQ_1(t) v_1,
\ee
and similarly for the second quadrature. The second point can be seen from \er{P0}, which can be used to relate classical and quantum expectations. The result is that if we define a system operator $\tilde V(t)$ acting on system state vectors $\psi$ to be the solution of the SDE
\be
d\tilde V(t)= \left\{ -K^\mu(u(t)) dt + L dq_1(t)
+ M dq_2(t)
\right\} \tilde V(t)
\label{tilde-V-qsde}
\ee
where we have used the components of the Wiener process $q(t)=(q_1(t), q_2(t))$ which describes the statistics of the real quadrature as mentioned above, then
for any system operator $X$
\begin{eqnarray}
\la \rho_0, V^\dagger(t) [ X\otimes I \otimes I ] V(t) \ra 
= \E^0[\la \pi_0, \tilde V^\dagger (t) X \tilde V(t) \ra ] ,
\label{stoch-rep-1}
\end{eqnarray}
where $\E^0$ denotes expectation with respect to the reference probability distribution $\bP^0$ (recall \er{P0}).
Therefore if we write
\be
\tilde j^\mu_t(u,X)= \tilde V^\dagger(t) X \tilde V(t) ,
\ee
we obtain
\be
J^\mu(\K) = \E^0[ \la \pi_0, \tilde j^\mu_T(u,e^{\mu C_2}) \ra ] ,
\label{cost-rep}
\ee
 a stochastic representation of the risk-sensitive cost function with respect to the reference Wiener distribution $\bP^0$ (in terms of semigroups  $\tilde\Phi^\mu_t(X)=\qE_{v_1\otimes v_2}[j^\mu_t(u,X)]=\E^0[\tilde j^\mu_t(u,X)]$).

Consider the operators $Y_2(t)$, $0 \leq t \leq T$ describing the real quadrature of the second output field channel (and containing information about the interaction with the system). These operators are also commutative, since from \er{bqnd-y},
\be
[Y_2(t),Y_2(s)]=0 \ \forall \ 0 \leq s, t \leq T .
\ee
Since we only measure the second field channel, we average the first component by computing the classical conditional expectation
\be
\hat j^\mu_t(u,X) = \E^0[ \tilde j^\mu_t(u,X) \vert q_2(s), 0 \leq s \leq t ]  .
\ee
This is straightforward due to the independence of the two fields; indeed, 
we note that $\tilde j^\mu_t(u,\cdot)$ satisfies, with respect to the reference distribution $\bP^0$, the SDE
\begin{eqnarray}
d \tilde j^\mu_t(u,X) &=& \tilde j^\mu_t(u,-XK^\mu(u(t))   -K^{\mu \dagger}(u(t))X
\nonumber \\
&& + L^\dagger XL + M^\dagger XM    )dt \nonumber
\\
&&+\tilde j^\mu_t(u, L^\dagger X + XL) dq_1(t) 
\nonumber \\
&&+ \tilde j^\mu_t(u, M^\dagger X + XM) dq_2(t) 
\label{tilde-j-sde} 
\end{eqnarray}
with corresponding output equation
\be
dy_2(t) = dq_2(t) 
\ee
(since the measurement values are $y_2(t)=q_2(t)$).
Then, say from classical filtering theory, \cite[Chapter 18]{RE82}, \cite[Chapter 7]{WH85},  we have
\begin{eqnarray}
d \hat j^\mu_t(u,X) &=& \hat j^\mu_t(u,-XK^\mu(u(t))   -K^{\mu \dagger}(u(t))X 
\nonumber \\
&&+ L^\dagger XL + M^\dagger XM    )dt 
 \nonumber \\
 &&+ \hat j^\mu_t(u, M^\dagger X + XM) dy_2(t) .
\label{hat-j-sde} 
\end{eqnarray}
By standard properties of classical conditional expectations (see, e.g. \cite[Section 34]{PB79}), we have
\begin{eqnarray*}
&&\E^0[ \la \pi_0, \tilde j^\mu_t(u,X) \ra ] 
\\
 &= &\E^0[\E^0[ \la \pi_0, \tilde j^\mu_t(u,X) \ra \vert y_2(s), 0 \leq s \leq t] ],
\end{eqnarray*}
and hence
\be
J^\mu(\K) = \E^0[ \la \pi_0, \hat j^\mu_T(u,e^{\mu C_2}) \ra ] . 
\label{J-rs-rep-0}
\ee
Here, the expectation is with respect to $y_2(t)$ ($=q_2(t)$), a standard Wiener process (since $q_1(t)$ has been averaged out in the definition of $\hat j^\mu_t(u,X)$).

We define an unnormalized {\em risk-sensitive state} $\sigma^\mu_t$ (which acts on system operators) by
\be
\la \pi_0, \hat j^\mu_t(u,X) \ra = \la \sigma^\mu_t, X \ra .
\ee
Then $\sigma^\mu_t$ is the solution of the SDE
\begin{eqnarray}
d \sigma^\mu_t &=&  (-K^\mu(u(t))\sigma^\mu_t   -\sigma^\mu_tK^{\mu \dagger}(u(t)) 
 \nonumber \\
 &&+ L\sigma^\mu_t L^\dagger  + M\sigma^\mu_t M^\dagger    )dt 
 \nonumber \\
 &&+ (M\sigma^\mu_t + \sigma^\mu_t M^\dagger ) dy_2(t) ,
\label{sigma-mu-sde}
\end{eqnarray}
or
\begin{eqnarray}
d \sigma^\mu_t &=&  -\frac{i}{\hbar}[ H(u(t)), \sigma^\mu_t] dt + \cD[L]\sigma^\mu_t dt+ \cD[M]\sigma^\mu_t dt
\nonumber \\
&&+  \frac{\mu}{2}\tilde{\cH}[C_1(u(t))]\sigma^\mu_t dt+   \tilde{\cH}[M]\sigma^\mu_t dy_2(t) ,
\label{sigma-mu-sde-a}
\end{eqnarray}
where $\cD[c]\rho = c\rho c^\dagger -\demi (c^\dagger c \rho + \rho c^\dagger c)$ and $\tilde{\cH}[c]\rho= c\rho + \rho c^\dagger$. Equation \er{sigma-mu-sde} (or \er{sigma-mu-sde-a}, or \er{tilde-pi-sde-a} below) is called the {\em risk-sensitive filter}.

The representation \er{J-rs-rep-0} becomes
\be
J^\mu(\K) = \E^0[ \la \sigma^\mu_T, e^{\mu C_2} \ra ] . 
\label{J-rs-rep-1}
\ee
This expression is similar to the classical forms \cite[eq. (3.4)]{BV85}, \cite[eq. (2.10)]{JBE94}, and will be used in Section \ref{sec:dp}.

Next we point out that the risk-sensitive state $\sigma^\mu_t$ reduces to the standard conditional state $\sigma_t$ when $\mu=0$, and \er{sigma-mu-sde} or \er{sigma-mu-sde-a} reduce to the usual stochastic master equation or Belavkin quantum filtering equation (e.g. \cite[Chapter 5.2.5]{AH01}); indeed
\begin{eqnarray}
d \sigma_t &=&  (-K(u(t))\sigma_t   -\sigma_tK^{\dagger}(u(t)) 
\nonumber \\
&&+ L\sigma_t L^\dagger  + M\sigma_t M^\dagger    )dt 
\nonumber \\
&& + (M\sigma_t + \sigma_t M^\dagger ) dy_2(t) .
\label{sigma-sde}
\end{eqnarray}
The corresponding normalized conditional state is given by
\be
\pi_t = \frac{\sigma_t}{\la \sigma_t,1 \ra} ,
\ee
which satisfies the SDE
\begin{eqnarray}
d \pi_t  &=&  (-K(u(t))\pi_t   -\pi_tK^{\dagger}(u(t)) 
\label{pi-sde} \\
&& + L\pi_t L^\dagger  + M\pi_t M^\dagger    )dt 
 \nonumber \\
 &&+ (M\pi_t + \pi_t M^\dagger    -\pi_t \tr [(M+M^\dagger)\pi_t]   ) dw(t) ,
\nonumber
\end{eqnarray}
where $w(t)$ is a standard Wiener process (innovation) under a distribution $\bP$ (to be described shortly) related to $y_2(t)$ by
\be
dy_2(t) = \tr[(M+M^\dagger)\pi_t] dt + dw(t) .
\label{yt}
\ee
Equivalently,
\begin{eqnarray}
d\pi_t &=& -\frac{i}{\hbar} [H(u(t)), \pi_t ] dt + \cD[L]\pi_t dt+ \cD[M]\pi_tdt
\nonumber \\
&&\hspace{0.5cm} + \cH[M]\pi_t dw(t) ,
\label{pi-sde-a}
\end{eqnarray}
where $\cH[c]\rho= c\rho + \rho c^\dagger - \rho\tr(c\rho + \rho c^\dagger)$.

The distribution $\bP$ is defined on the set $\Omega_{2,T}$ of all possible measurement  paths $y_2(t)$, $0 \leq t \leq T$ of the operators $Y_2(s)$, $0 \leq s \leq T$. For a subset  $F_2 \subset \Omega_{2,T}$, the associated projection operator is denoted $P^{Y_2}_T(F)$, and the corresponding probability is given by
\begin{eqnarray}
\bP(F_2) & = & \la \rho_0, P^{Y_2}(F_2) \ra \nonumber
\\
& = & \la \rho(T), P^{Q_2}_{T}(F_2) \ra ,
\label{output-prob}
\end{eqnarray}
where $\rho(t) = U(t)  \rho_0 U^\dagger(t)$, $\rho_0=\pi_0 \otimes v_1v_1^\dagger \otimes v_2v_2^\dagger$. Note that the distribution $\bP$ depends on the controller $\K$ in the feedback loop, so strictly $\bP=\bP^\K$.

We conclude this section with some alternative expressions for the risk-sensitive cost together with the associated variants of the risk-sensitive state. These can be derived using the stochastic calculus, e.g. \cite{RE82}, \cite{WH85}.
If we define a second unnormalized risk-sensitive state by
\be
\bar \pi^\mu_t = \frac{\sigma^\mu_t}{\la \sigma_t,1 \ra} 
\ee
(note that the denominator   is the normalization factor for the standard conditional state), we obtain the representation
\be
J^\mu(\K) = \E[ \la \bar \pi^\mu_T, e^{\mu C_2} \ra ] 
\label{J-rs-rep-2}
\ee
with respect to the output distribution $\bP$ (since $d\bP=\la \sigma_T,1\ra d\bP^0$, \cite[Chapter 5]{AH01}, \cite[Section 32]{PB79}, \cite[Chapters 6 and 7]{WH85}). A third representation can be obtained using the following normalized risk-sensitive state
\be
\pi^\mu_t = \frac{\sigma^\mu_t}{\la \sigma^\mu_t,1 \ra} ,
\ee
namely 
\be
J^\mu(\K) = \E^\mu[    \exp(\mu  \int_0^T \tr(C_1(u(t) ) \pi^\mu_t   ) dt)
    \la \pi^\mu_T, e^{\mu C_2} \ra ] 
\label{J-rs-rep-3}
\ee
where $\E^\mu$ denotes expectation with respect to the probability distribution $\bP^\mu$ defined by
$d\bP^\mu=\Lambda^\mu_T d\bP^0$, where
\begin{eqnarray*}
\Lambda^\mu_T &=& \exp(  -\demi \int_0^T \vert  \tr[ (M+M^\dagger)\pi^\mu_t  ]  \vert^2 dt  
\\
&& +
\int_0^T   \tr[ (M+M^\dagger)\tilde\pi^\mu_t  ]    dy_2(t) ) .
\end{eqnarray*}
The SDE satisfied by $\pi^\mu_t$ is
\begin{eqnarray}
d\pi^\mu_t &=& -\frac{i}{\hbar} [H(u(t)), \pi^\mu_t ] dt + \cD[L]\pi^\mu_t dt+ \cD[M]\pi^\mu_tdt \nonumber \\
&&+ \frac{\mu}{2} \cH[C_1(u(t)) ] \pi^\mu_t dt
+ \cH[M]\pi^\mu_t dw^\mu(t) ,
\label{tilde-pi-sde-a}
\end{eqnarray}
where $w^\mu(t)$ is a standard Wiener process with respect to $\bP^\mu$ defined by
\be
dy_2(t) = \tr[(M+M^\dagger)\pi^\mu_t] dt + dw^\mu(t) .
\label{yt-a}
\ee

In the next section we use $\sigma^\mu_t$ and the representation \er{J-rs-rep-1} to show how dynamic programming methods from optimal control theory can be applied to the risk-sensitive problem.

The state $\pi_t$ (or equivalently $\sigma_t$) is the familiar a posteriori state of the system conditioned on the measurement data, whose function is to facilitate the {\em description} of behavior via calculation of conditional expected values of observable quantities, and thus is a representation of measurement knowledge. See \cite[Section IV]{WM93b} for an interesting discussion of the reality of quantum trajectories and conditional states. The risk-sensitive state $\pi^\mu_t$ (or equivalently $\sigma^\mu_t$, $\tilde\pi^\mu_t$)  is also  determined by the measurement data, but through dynamics that contains the cost term $C_1$---it is by this mechanism that knowledge is tempered by purpose, and is a reflection of the {\em prescriptive} nature of control theory.  The risk-sensitive state is properly understood in the context of the risk-sensitive feedback control problem, since it is a suitable state in terms of which the risk-sensitive problem can be solved. This is a different use of measurement information than is standard in quantum mechanics (see \cite{RWB76} for a short comparison of descriptive and prescriptive sciences). All of these states are examples of what are called {\em information states} in classical control theory, \cite{KV86}, \cite{JBE94}. 

We remark that the quantum formulation of the risk-sensitive problem given here corresponds in discrete time to \cite[case (i) of Example 6]{J04}. The state $\bar\pi^\mu_t$ was used (there denoted $\hat\omega_k$), and the modified stochastic master equation in discrete time is \cite[eq. (39)]{J04}.

We close this section by noting that if the field $Y_2(t)$ is measured with efficiency $0 \leq \eta \leq 1$, that is, if we measure
\be
Z(t)= \sqrt{\eta} \, Y_2(t) + \sqrt{1-\eta} \, Y_3(t)
\ee
instead of $Y_2(t)$, where $Y_3(t)$ is the real quadrature of a third (and independent)  field, then the risk-sensitive filter equation \er{sigma-mu-sde} becomes
\begin{eqnarray}
d \sigma^\mu_t &=&  (-K^\mu(u(t))\sigma^\mu_t   -\sigma^\mu_tK^{\mu \dagger}(u(t)) 
 \nonumber \\
 &&+ L\sigma^\mu_t L^\dagger  + M\sigma^\mu_t M^\dagger    )dt 
 \nonumber \\
 &&+ \sqrt{\eta} (M\sigma^\mu_t + \sigma^\mu_t M^\dagger ) dz(t) .
\label{sigma-mu-sde-eta}
\end{eqnarray}

\section{Dynamic Programming}
\label{sec:dp}

In this section we show how the dynamic programming method can be used to determine the optimal controller. We make use of the representation \er{J-rs-rep-1} and the fact that the state $\sigma^\mu_t$ evolves in time according to the dynamics \er{sigma-mu-sde} driven by the measurement data $y(\cdot)$.

The method of dynamic programming works by defining, for each time $t$ and state $\sigma$, the optimal value of the cost from time $t$ to the final time $T$. This optimal value is called the {\em value function}. By considering the equation that the value function satisfies, called the {\em dynamic programming equation} or {\em Hamilton-Jacobi-Bellman equation}, the optimal control that should be used at time $t$ when in state $\sigma$ can be determined. The dynamic programming equation for the value function is solved backwards, from the final time $t=T$ to the initial time $t=0$. The equation is an infinitesimal statement of the {\em principle of optimality},  \cite[Chapter VI]{FR75}, \cite[Chapter 6]{KV86}.

Define the risk-sensitive value function $S^\mu(\sigma,t)$ for an arbitrary initial unnormalized state $\sigma$ and initial time $0 \leq t \leq T$ by
\be
S^\mu(\sigma,t) = \inf_{\K}  \E^0_{\sigma,t}[\la  \sigma^\mu_T, e^{\mu C_2} \ra ] ,
\label{rs-value-def}
\ee
where $\sigma^\mu_T$ denotes the solution of \er{sigma-mu-sde} at time $T$ with initial condition $\sigma^\mu_t=\sigma$ (we have made explicit the dependence on the initial state and time in the expectation notation). Note that the cost \er{J-rs} is given by
\be
J^\mu(\K) =  \E^0_{\pi_0,0}[\la  \sigma^\mu_T, e^{\mu C_2} \ra ]  
\ee
so that the optimal controller $\K^{\mu,\star}$ is determined by
\be
J^\mu(\K^{\mu,\star}) = S^\mu(\pi_0,0) .
\ee

The method of dynamic programming in this context relates the value function at time $t$ and at a later time $t \leq s \leq T$ along optimal trajectories via the relation
\be
 S^\mu(\sigma,t) = \inf_{\K} \E^0_{\sigma,t}[S^\mu(\sigma^\mu_s,s) ]   .
\label{rs-dp}
\ee
This is the principle of optimality.
Note that by definition the terminal value is $S^\mu(\sigma,T)=\la \sigma, e^{\mu C_2} \ra$. The dynamic programming relation may be considered in differential form (subject to mathematical technicalities) resulting in the dynamic programming PDE (see \er{rs-dpe} below). To see this, let $h > 0$, set $s=t+h$ in \er{rs-dp}, re-arrange and divide by $h$:
\be
0 =  \inf_{\K} \E^0[ \frac{S^\mu(\sigma^\mu_{t+h}, t+h) - S^\mu(\sigma,t)}{h}  ]  .
\ee
Sending $h \downarrow 0$ yields 
\be
\ba{rl}
\frac{\partial }{\partial t} S^\mu(\sigma,t) + \ds{\inf_{u\in\bU}}
\cL^{\mu;u} S^\mu(\sigma,t) & = 0, \  \ 0 \leq t < T,
\\
S^\mu(\sigma,T) & = \la \sigma, e^{\mu C_2} \ra .
\ea
\label{rs-dpe}
\ee
Note that in the dynamic programming PDE \er{rs-dpe}, the minimization is over the control values $u$, whereas in the definitions of the cost \er{J-rs} and value function \er{rs-value-def} the minimizations are over the controllers $\K$.

We now explain the meaning of the operator
  $\cL^{\mu,u}$ appearing in the dynamic programming PDE \er{rs-dpe}, following  classical stochastic control methods \cite{F82}.     For a fixed constant control value $u$ (i.e. $u(t)=u\in\bU$ for all $t$), $\sigma^\mu_t$ is a Markov process with generator   $\cL^{\mu,u}$, which is defined, when it
exists, by
\be
\cL^{\mu,u} f(\sigma) = \lim_{t \downarrow 0}
\frac{ \E^0_{\sigma,0}[ f(\sigma^\mu_t)] - f(\sigma)}{t} 
\label{rs-gen-def}
\ee
for suitably smooth functions $f(\cdot)$. In fact, $\cL^{\mu,u} f(\sigma) $ can be calculated explicitly for $f$ of the form
\be
 f(\sigma) = g( \la \sigma, X_1 \ra, \ldots,
\la \sigma, X_J \ra ),
\label{f}
\ee
where $g$ is a smooth bounded function of vectors of length $J$, and $X_1,\ldots,X_J  $ are system operators. Indeed,  for fixed $u \in\bU$ and  for such $f$, we have
\begin{eqnarray}
&&\cL^{\mu; u} f(\sigma) =
\label{rs-gen} \\
 &&\frac{1}{2} \sum_{j,k=1}^J
g_{jk}( \la  \sigma, X_1 \ra, \ldots,
\la  \sigma, X_J \ra ).
\nonumber \\
&& .\la  \sigma, M^\dagger X_j + X_j M \ra 
\la \sigma, M^\dagger X_k + X_k M \ra
\nonumber \\
&& + \sum_{j=1}^J g_j ( \la \sigma, X_1 \ra,
\ldots,
\la  \sigma, X_J \ra ).
\nonumber \\
&& .
\la \sigma, - K^\mu(u)X_j - X_j K^\mu(u) +L^\dagger X_j L + M^\dagger X_j M  \ra
\nonumber
\end{eqnarray}
where $g_j$ and $g_{jk}$ denote first and second order partial
derivatives of $g$.

If the dynamic programming PDE has a sufficiently smooth solution $S^\mu(\sigma,t)$, then the optimal controller $\K^{\mu,\star}$ can be obtained as follows. Let $\bu^{\mu,\star}(\sigma,t)$ denote the control value that attains the minimum in \er{rs-dpe} for each $\sigma$, $t$. The optimal controller is obtained by combining this function with the risk-sensitive filter  \er{sigma-mu-sde}:
\be
\K^{\mu,\star} \ : \ \ba{rl}
d \sigma^\mu_t & =  (-K^\mu(u(t))\sigma^\mu_t   -\sigma^\mu_tK^{\mu \dagger}(u(t)) 
\\
& \hspace{0.5cm} + L\sigma^\mu_t L^\dagger  + M\sigma^\mu_t M^\dagger    )dt 
 \\
& \hspace{0.5cm} + (M\sigma^\mu_t + \sigma^\mu_t M^\dagger ) dy_2(t) 
\\
u(t) & = \bu^{\mu,\star}(\sigma^\mu_t,t) .
\ea
\label{K-star}
\ee

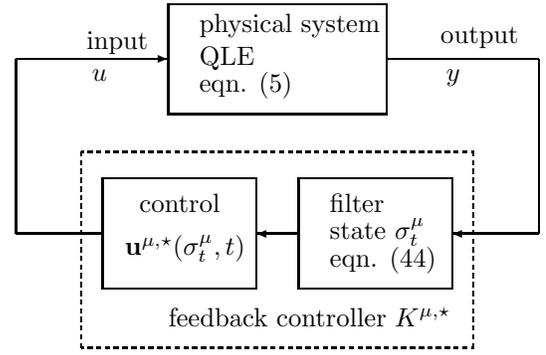
\begin{figure}[htb]
\begin{center}
\setlength{\unitlength}{1800sp}%
\begingroup\makeatletter\ifx\SetFigFont\undefined%
\gdef\SetFigFont#1#2#3#4#5{%
  \reset@font\fontsize{#1}{#2pt}%
  \fontfamily{#3}\fontseries{#4}\fontshape{#5}%
  \selectfont}%
\fi\endgroup%
\begin{picture}(7224,4749)(889,-4798)
\thinlines
{\color[rgb]{0,0,0}\put(3001,-1561){\framebox(3000,1500){}}
}%
{\color[rgb]{0,0,0}\put(2101,-3961){\framebox(2100,1500){}}
}%
{\color[rgb]{0,0,0}\put(4801,-3961){\framebox(2100,1500){}}
}%
{\color[rgb]{0,0,0}\put(4801,-3211){\vector(-1, 0){600}}
}%
{\color[rgb]{0,0,0}\put(2101,-3211){\line(-1, 0){1200}}
\put(901,-3211){\line( 0, 1){2400}}
\put(901,-811){\vector( 1, 0){2100}}
}%
{\color[rgb]{0,0,0}\put(6001,-811){\line( 1, 0){2100}}
\put(8101,-811){\line( 0,-1){2400}}
\put(8101,-3211){\vector(-1, 0){1200}}
}%
{\color[rgb]{0,0,0}\put(1801,-4786){\dashbox{60}(5400,2700){}}
}%
\put(1951,-1111){\makebox(0,0)[lb]{\smash{\SetFigFont{10}{14.4}{\familydefault}{\mddefault}{\updefault}{\color[rgb]{0,0,0}$u$}%
}}}
\put(6826,-1111){\makebox(0,0)[lb]{\smash{\SetFigFont{10}{14.4}{\familydefault}{\mddefault}{\updefault}{\color[rgb]{0,0,0}$y$}%
}}}
\put(1876,-661){\makebox(0,0)[lb]{\smash{\SetFigFont{10}{14.4}{\familydefault}{\mddefault}{\updefault}{\color[rgb]{0,0,0}input}%
}}}
\put(6751,-586){\makebox(0,0)[lb]{\smash{\SetFigFont{10}{14.4}{\familydefault}{\mddefault}{\updefault}{\color[rgb]{0,0,0}output}%
}}}
\put(3426,-1261){\makebox(0,0)[lb]{\smash{\SetFigFont{10}{14.4}{\familydefault}{\mddefault}{\updefault}{\color[rgb]{0,0,0}eqn. \er{X-qle}}%
}}}
\put(3426,-886){\makebox(0,0)[lb]{\smash{\SetFigFont{10}{14.4}{\familydefault}{\mddefault}{\updefault}{\color[rgb]{0,0,0}QLE}%
}}}
\put(3426,-436){\makebox(0,0)[lb]{\smash{\SetFigFont{10}{14.4}{\familydefault}{\mddefault}{\updefault}{\color[rgb]{0,0,0}physical system}%
}}}
\put(2601,-2911){\makebox(0,0)[lb]{\smash{\SetFigFont{10}{14.4}{\familydefault}{\mddefault}{\updefault}{\color[rgb]{0,0,0}control}%
}}}
\put(2401,-3511){\makebox(0,0)[lb]{\smash{\SetFigFont{10}{14.4}{\familydefault}{\mddefault}{\updefault}{\color[rgb]{0,0,0}$\bu^{\mu,\star}(\sigma^\mu_t,t)$}%
}}}
\put(5226,-2911){\makebox(0,0)[lb]{\smash{\SetFigFont{10}{14.4}{\familydefault}{\mddefault}{\updefault}{\color[rgb]{0,0,0}filter}%
}}}
\put(5226,-3286){\makebox(0,0)[lb]{\smash{\SetFigFont{10}{14.4}{\familydefault}{\mddefault}{\updefault}{\color[rgb]{0,0,0}state $\sigma^\mu_t$}%
}}}
\put(5226,-3661){\makebox(0,0)[lb]{\smash{\SetFigFont{10}{14.4}{\familydefault}{\mddefault}{\updefault}{\color[rgb]{0,0,0}eqn. \er{sigma-mu-sde}}%
}}}
\put(3001,-4486){\makebox(0,0)[lb]{\smash{\SetFigFont{10}{14.4}{\familydefault}{\mddefault}{\updefault}{\color[rgb]{0,0,0}feedback controller $K^{\mu,\star}$}%
}}}
\end{picture}

\caption{Optimal risk-sensitive controller $K^{\mu,\star}$ in closed loop with the physical system being controlled.}
\label{fig:fb-rs}
\end{center}
\end{figure}

This controller is a dynamical controller, of the general form \er{K-dyn}. 
The structure of this controller is illustrated in Figure \ref{fig:fb-rs}, where it is shown in closed loop with the quantum system being controlled. The controller $K^{\mu,\star}$ can be implemented in classical electronics (e.g. analog circuit or digital computer). The filter is the implementation of the dynamics \er{sigma-mu-sde} for the risk-sensitive state $\sigma^\mu_t$, which as described at the end of Section \ref{sec:rep} represents the controller's knowledge of the physical system tempered by its purpose. The feedback control function $\bu^{\mu, \star}(\sigma,t)$ is determined by solving the dynamic programming equation backwards; this computation can be done offline, with the results stored and available for online use. The risk-sensitive filter is, of course, solved online while the quantum system is being controlled.

Note that we may, of course, alternatively carry out dynamic programming and express the controller in terms of the states $\pi^\mu_t$, $\tilde \pi^\mu_t$; in fact, appropriate normalization is important for practical reasons.

\section{Risk-Neutral Optimal Control}
\label{sec:rn}

In this section we briefly discuss a risk-neutral problem of the type that has been studied by \cite{VPB83}, \cite{VPB88}, \cite{DJ99}, \cite{BEB04}. Specifically, we consider the {\em risk-neutral} problem defined by the quantum
expectation
\be
J(\K) = \la \rho_0, 
\int_0^T C_1(t) dt + C_2(T)
\ra ,
\label{J-rn}
\ee
where as before, $\rho_0 =  \pi_0 \otimes \vacv_1\vacv^\dagger_1 \otimes
\vacv_2\vacv^\dagger_2$. The key step in solving the optimal control problem specified by \er{J-rn} is again a stochastic representation followed by classical conditional expectation, as in Section \ref{sec:rep}, which results in
\begin{eqnarray}
J(\K) &=& \E[ \int_0^T \la \pi_t, C_1(u(t)) \ra dt + \la \pi_T, C_2 \ra ]
\nonumber \\
 &=& \E^0[ \int_0^T \la \sigma_t, C_1(u(t)) \ra dt + \la \sigma_T, C_2 \ra ]
\label{J-rn-rep-0}
\end{eqnarray}
where $\pi_t$ and  $\sigma_t$  are the conditional states, assuming interchanges of expectations and integrals are justified.

The risk-neutral value function can now be defined by
\be
W(\sigma,t) = \inf_{\K} \E^0_{\sigma,t} [ \int_t^T \la \sigma_s, C_1 \ra ds + \la \sigma_T, C_2 \ra   ]
\label{rn-value}
\ee
and the corresponding dynamic programming equation reads
\be
\ba{rl}
\frac{\partial }{\partial t} W(\sigma,t) + \ds{\inf_{u\in\bU}} \{
\cL^{u} W(\sigma,t)  + C_1(u)   \}& = 0, \  \ 0 \leq t < T,
\\
W(\sigma,T) & = \la \sigma, C_2 \ra .
\ea
\label{rn-dpe}
\ee
where, for fixed control value $u\in\bU$, $\cL^{u}$ is the generator of the Markov process $\sigma_t$, and is given by
\begin{eqnarray}
&&\cL^{u} f(\sigma) =
\label{rn-gen} \\
 &&\frac{1}{2} \sum_{j,k=1}^J
g_{jk}( \la  \sigma, X_1 \ra, \ldots,
\la  \sigma, X_J \ra ).
\nonumber \\
&& .\la  \sigma, M^\dagger X_j + X_j M \ra 
\la \sigma, M^\dagger X_k + X_k M \ra
\nonumber \\
&& + \sum_{j=1}^J g_j ( \la \sigma, X_1 \ra,
\ldots,
\la  \sigma, X_J \ra ).
\nonumber \\
&& .
\la \sigma, - K(u)X_j - X_j K(u) +L^\dagger X_j L + M^\dagger X_j M  \ra
\nonumber
\end{eqnarray}
for functions $f$ of the form \er{f}.

If the dynamic programming equation \er{rn-dpe}  has a sufficiently smooth solution $W(\sigma,t)$, then the optimal controller $\K^\star$ is given by
\be
\K^{\star} \ : \ \ba{rl}
d \sigma_t & =  (-K(u(t))\sigma_t   -\sigma_tK^{\dagger}(u(t)) 
\\
& \hspace{0.5cm} + L\sigma_t L^\dagger  + M\sigma_t M^\dagger    )dt 
 \\
& \hspace{0.5cm} 
+ (M\sigma_t + \sigma_t M^\dagger ) dy_2(t) 
\\
u(t) & = \bu^{\star}(\sigma^\mu_t,t) .
\ea
\label{K-star-rn}
\ee
where $\bu^\star(\sigma,t)$ attains the minimum in \er{rn-dpe}. The dynamical part of this controller is the Belavkin quantum filter, \er{sigma-sde}.

\section{Feedback Control of a Two-Level Atom}
\label{sec:2l}

In this section we consider the application of the risk-sensitive control problem to the example studied in \cite{BEB04}, namely the feedback control of a two-level atom using a laser.

The amplitude and phase of the input laser can be adjusted, so via the interaction with the laser  the atom can be controlled. The real quadrature of a second field channel is continuously monitored, say by homodyne detection, providing an indirect measurement of  the atom. The control input is complex, $u=u_r + i u_i=\vert u \vert e^{i \text{arg}u}\in\C$ (the control field channel becomes a coherent state corresponding to $u$). The measurement signal $y_2(t)$ is real. It is desired to regulate the system in the $\sigma_z$ up state $\vert \up \ra = (1,0)^T$ (the down state is $\vert \dn \ra = (0,1)^T$, and
 $\sigma_x=\left( \ba{cc} 0&1\\ 1&0 \ea \right)$, $\sigma_y=\left( \ba{cc} 0&-i\\ i&0 \ea \right)$ and $\sigma_z=\left( \ba{cc} 1&0\\ 0&-1 \ea \right)$ are the Pauli matrices).

In terms of the notation used in this paper, we have
\begin{eqnarray*}
L&=& \kappa_f \sigma_-, \  \ M= \kappa_s \sigma_-, \  \  H(u)=i (u^\star L - u L^\dagger), \ \ \\ &&\kappa_f^2+\kappa_s^2=1, \ \ 
\sigma_- = \left(   \ba{cc} 0 & 0 \\ 1 & 0 \ea \right), 
\\
\bU&=&\C, \ \  C_1(u)= a \left(   \ba{cc} 0 & 0 \\ 0 & 1 \ea \right)  + \demi b \vert u \vert^2  \left(   \ba{cc} 1 & 0 \\ 0 & 1 \ea \right),  \ \ 
\\
&& C_2=  c \left(   \ba{cc} 0 & 0 \\ 0 & 1 \ea \right) , \ \ a \geq 0, b \geq 0, c\geq 0.
\end{eqnarray*}
Here, $\kappa_f^2$ and $\kappa_s^2$ are the decay rates into the control and measurement channels. The parameters $a$, $b$ and $c$ are weights for the components of the cost. Note that $\la \dn \vert C_1(u) \vert \dn \ra > 0$ and $\la \dn \vert C_2 \vert \dn \ra > 0$ (if $a>0$ and $c>0$), while $\la \up \vert C_1(0) \vert \up \ra = 0$ and $\la \up \vert C_2 \vert \up \ra =0$, reflecting the control objective.

We use the framework described in previous sections to solve the optimal risk-sensitive control problem. Since the second ($u$-dependent) part of $C_1(u)$ is proportional to the identity, that part commutes  with all operators and it is convenient to factor its contribution to the risk-sensitive state  by writing
\begin{eqnarray*}
\sigma^\mu_t & =&   \demi\left( n(t)I + x(t) \sigma_x + y(t) \sigma_y + z(t) \sigma_z\right) .
\\ &&  \hspace{0.5cm} .\exp\left(  \demi \mu b \int_0^t \vert u(s)\vert^2 ds    \right) 
\\
& =&    \demi \left( \ba{cc}  n(t) + z(t)  & x(t)-iy(t)  \\    x(t)+iy(t)   &  n(t)-z(t)  \ea  \right).
\\ &&  \hspace{0.5cm} .
\exp\left(  \demi \mu b\int_0^t \vert u(s)\vert^2 ds    \right)  .
\end{eqnarray*}
Then substitution into the SDE \er{sigma-mu-sde} shows that the coefficients satisfy the SDEs 
\begin{eqnarray}
d n(t) &=&  \demi \mu a (n(t)-z(t)) dt + \kappa_s x(t) dy_2(t)
\label{px-sde}\\
dx(t) &=& -\demi(1-\mu a) x(t)dt +2\kappa_f u_r(t) z(t)dt 
\nonumber \\
&&+ \kappa_s(n(t)+z(t))dy_2(t)
\nonumber \\
dy(t) &=& -\demi(1-\mu a) y(t)dt -2\kappa_f u_i(t) z(t)dt 
\nonumber \\
dz(t) &=&-(1-\demi \mu a)z(t)dt -(1+\demi\mu a)n(t)dt
\nonumber \\
&& -2\kappa_f (u_r(t) x(t) - u_i(t)y(t))dt 
 -\kappa_s x(t)dy_2(t) .
\nonumber 
\end{eqnarray}
The representation \er{J-rs-rep-1} reads
\be
J^\mu(\K) = \E^0[ \exp\left(  \demi \mu \int_0^T b\vert u(s)\vert^2 ds    \right) 
\demi (n(T)-z(T))e^{\mu c} ] .
\label{J-rs-rep-1-2l}
\ee
We consider the value function \er{rs-value-def} as a function of the coefficients, i.e. $S^\mu(n,x,y,z,t)$. In terms of these parameters, the dynamic programming equation is
\be
\ba{l}
\frac{\partial }{\partial t} S^\mu(n,x,y,z,t) + \ds{\inf_{u\in\C}} \{
\cL^{\mu;u} S^\mu(n,x,y,z,t)  
\\ \hspace{0.5cm} +   \demi \mu b\vert u \vert^2 S^\mu(n,x,y,z,t)     \} = 0, \  \ 0 \leq t < T,
\\
S^\mu(n,x,y,z,T)  = \demi (n-z)e^{\mu c} ,
\ea
\label{rs-dpe-2l}
\ee
where the operator $\cL^{\mu;u}$ is given, for sufficiently smooth functions $f(n,z,y,z)$, by
\begin{eqnarray*}
\cL^{\mu;u} f & =&  \demi \kappa_s^2 x^2 f_{nn} +\demi \kappa_s^2 (n+z)^2 f_{xx} +\demi \kappa_s^2 x^2 f_{zz}
\\
&& + \kappa_s^2 x (n+z) f_{nx} -\kappa_s^2 x^2 f_{nz} - \kappa_s^2 (n+z)xf_{xz}
\\
&& + f_n (\demi \mu a(n-z)) +f_x (-\demi(1-\mu a) x +2\kappa_f u_r z) 
\\
&&+ f_y (-\demi(1-\mu a) y -2\kappa_f u_i z ) 
\\
&& + f_z (-(1-\demi \mu a)z -(1+\demi\mu a)n 
\\&&-2\kappa_f (u_r x - u_i y) .
\end{eqnarray*}
Here, the subscripts $f_{nx}$, etc,  refer to partial derivatives, and the arguments $n,x,y,z$ have been omitted.

To construct the optimal risk-sensitive controller $\K^{\mu, \star}$, we suppose that \er{rs-dpe-2l} has a smooth solution, which we write as
\be
S^\mu(n,x,y,z,t)= n \exp \left(  \frac{\mu}{n} W^\mu(n,x,y,z,t)   \right) .
\ee
The minimum over $u$ in \er{rs-dpe-2l} can be explicitly evaluated by setting the derivatives of the expression in the parentheses (it is concave) with respect to $u_r$ and $u_i$ to zero. The result is
\begin{eqnarray}
\bu^{\mu,\star}_r(n,x,y,z,t)&=& \frac{2\kappa_f}{bn} ( x W^\mu_z(n,x,y,z,t)
\nonumber \\&& \hspace{0.5cm} -z W^\mu_x(n,x,y,z,t)  )
\nonumber \\
\bu^{\mu,\star}_i(n,x,y,z,t)&=& \frac{2\kappa_f}{bn} ( z W^\mu_y(n,x,y,z,t)
\nonumber \\&& \hspace{0.5cm} -y W^\mu_z(n,x,y,z,t)  ) .
\label{rs-u-star-2l}
\end{eqnarray}
The optimal risk-sensitive controller is then
\begin{eqnarray}
\K^{\mu,\star} \ : \ u(t) &=& \bu^{\mu,\star}_r(n(t),x(t),y(t),z(t),t) 
\nonumber 
\\&& \hspace{0.1cm} + i \bu^{\mu,\star}_i(n(t),x(t),y(t),z(t),t),
\label{rs-2l-k-star}
\end{eqnarray}
where $n(t)$, $x(t)$, $y(t)$ and $z(t)$ are given by \er{px-sde}.

Note that the dynamic programming equation \er{rs-dpe-2l} (which is a partial differential equation of parabolic type) is solved backwards in time, using the terminal condition specified: $S^\mu(n,x,y,z,T)  = \demi (n-z)e^{\mu c}$. The infimum in \er{rs-dpe-2l} can be removed by substituting in the optimal control values given by the explicit formulas \er{rs-u-star-2l}, if desired. However, the form \er{rs-dpe-2l} is better suited to numerical computation, since the optimal control structure is preserved, \cite{KD92}. Note that in this example, the risk-sensitive filter \er{sigma-mu-sde} is replaced by the finite-dimensional SDE \er{px-sde}; this fact is important for practical computational reasons.


Finally, we consider the risk-neutral problem. Write
\be
\sigma_t  =   \demi\left( n(t)I + x(t) \sigma_x + y(t) \sigma_y + z(t) \sigma_z\right) .
\ee
Then from the SDE \er{sigma-sde}, we find that
\begin{eqnarray}
d n(t) &=&   \kappa_s x(t) dy_2(t)
 \label{px-sde-rn}   \\
dx(t) &=& -\demi x(t)dt +2\kappa_f u_r(t) z(t)dt 
\nonumber \\
&& + \kappa_s(n(t)+z(t))dy_2(t)
\nonumber \\
dy(t) &=& -\demi y(t)dt -2\kappa_f u_i(t) z(t)dt 
\nonumber \\
dz(t) &=&-z(t)dt - n(t)dt 
\nonumber \\
&&-2\kappa_f (u_r(t) x(t) - u_i(t)y(t))dt 
-\kappa_s x(t)dy_2(t) .
\nonumber
\end{eqnarray}
The risk-neutral representation \er{J-rn-rep-0} becomes
\begin{eqnarray}
J(\K) &=& \E^0[\demi \int_0^T (a(n(t)-z(t)+b \vert u(t) \vert^2) dt 
\nonumber \\
&& \hspace{0.5cm} + \demi (n(T)-z(T))c ] ,
\label{J-rn-rep-0-2l}
\end{eqnarray}
and the dynamic programming equation is
\be
\ba{l}
\frac{\partial }{\partial t} W(n,x,y,z,t) + \ds{\inf_{u\in\C}} \{
\cL^{u} W(n,x,y,z,t)   
\\
+   \demi (a(n-z)+ b \vert u \vert^2 )     \} = 0, \  \ 0 \leq t < T,
\\
W(n,x,y,z,T)  = \demi (n-z)e^{c} ,
\ea
\label{rn-dpe-2l}
\ee
where
\begin{eqnarray*}
\cL^{u} f & =&  \demi \kappa_s^2 x^2 f_{nn} +\demi \kappa_s^2 (n+z)^2 f_{xx} +\demi \kappa_s^2 x^2 f_{zz}
\\
&& + \kappa_s^2 x (n+z) f_{nx} -\kappa_s^2 x^2 f_{nz} - \kappa_s^2 (n+z)xf_{xz}
\\
&& + f_x (-\demi x +2\kappa_f u_r z) 
\\
&&+ f_y (-\demi  y -2\kappa_f u_i z ) 
\\&&+ f_z (-z -n -2\kappa_f (u_r x - u_i y) 
\end{eqnarray*}
Evaluating the minimum in \er{rn-dpe-2l} gives
\begin{eqnarray}
\bu^{\star}_r(n,x,y,z,t)&=& \frac{2\kappa_f}{bn} ( x W_z(n,x,y,z,t)
\nonumber \\
&& \hspace{0.5cm} -z W_x(n,x,y,z,t)  )
\nonumber \\
\bu^{\star}_i(n,x,y,z,t)&=& \frac{2\kappa_f}{bn} ( z W_y(n,x,y,z,t)
\nonumber \\
&& \hspace{0.5cm} -y W_z(n,x,y,z,t)  ) ,
\label{rn-u-star-2l}
\end{eqnarray}
cf. \cite[eq. (15)]{BEB04}.
The optimal risk-neutral controller is 
\begin{eqnarray}
\K^{\star} \ : \ u(t) &=& \bu^{\star}_r(n(t),x(t),y(t),z(t),t) 
\nonumber \\
&& \hspace{0.5cm} + i \bu^{\star}_i(n(t),x(t),y(t),z(t),t),
\label{rn-2l-k-star}
\end{eqnarray}
where $n(t)$, $x(t)$, $y(t)$ and $z(t)$ are given by \er{px-sde-rn}. Note that normalization of \er{px-sde-rn} results in \cite[eq. (7)]{BEB04}.

Note that
the expressions for the both the risk-sensitive and risk-neutral controllers are similar, and involve a similar level of complexity for implementation.
When $a=0$, the risk-sensitive SDEs \er{px-sde} reduces to the risk-neutral or standard SDEs \er{px-sde-rn}, though the controllers will be different in general.

\section{Conclusion}
\label{sec:c}

In this paper we have studied a risk-sensitive optimal control problem for open quantum systems. The model we used for continuously monitored open quantum systems is given by a quantum Langevin equation.  Using quantum stochastic calculus and dynamic programming, we showed how to formulate and solve the risk-sensitive optimal control problem. The optimal controller we obtained is shown in Figure \ref{fig:fb-rs}. It has two components. One component is dynamic, a filter that computes the risk-sensitive state. The second component is an optimal control feedback function that is found by solving the dynamic programming equation. The optimal controller can be implemented using classical electronics. This procedure is computationally intensive, due to the storage requirements of the feedback control function, and the speed demands of online solution of the risk-sensitive filtering equations.

A significant feature of the optimal control solution is the use of the risk-sensitive state. This is different in general to the state usually used in quantum physics. The difference is because the filter that computes it contains terms corresponding to the cost function specifying the control objective. Such cost terms do not appear in the conventional quantum trajectory equations or Belavkin quantum filtering equations.
One could say that the risk-sensitive state describes  knowledge of the physical system being controlled, but tempered by the control purpose, and thereby represents measurement information in a way that is suitable for this feedback  control problem. Consideration of this issue may be of interest.

One of the motivations for considering risk-sensitive optimal control is the enhanced robustness properties relative to risk-neutral (e.g. LQG) control. Robustness of a control system concerns its ability to cope with performance degrading influences of uncertainty and noise. In the case of quantum systems, decoherence is a key limiting factor in the development of quantum technologies, and it is therefore important to design controllers which are also robust with respect to decoherence. Consequently, evaluation of the robustness properties of risk-sensitive control for open quantum systems is an important topic of investigation. 

We also mention that for quantum systems with quadratic Hamiltonians and Gaussian initial states, the risk-sensitive state is also Gaussian. This fact has important practical implications which are beginning to be investigated, \cite{WDDJ05}.


\noindent {\bf Acknowledgement.} The author wishes to thank A.~Doherty and L.~Bouten for helpful discussions.


\bibliographystyle{plain}


\end{document}